\documentclass[conference]{IEEEtran}
\IEEEoverridecommandlockouts

\usepackage{cite}
\usepackage{amsmath,amssymb,amsfonts}
\usepackage{algorithmic}
\usepackage{graphicx}
\usepackage{textcomp}
\usepackage{xcolor}
\usepackage{balance}
\usepackage{tikz}
\usetikzlibrary{arrows.meta,positioning}
\usepackage{pifont}
\usepackage{enumitem}
\usepackage{multirow}
\usepackage{booktabs}
\usepackage{url}
\usepackage{makecell}
\def\BibTeX{{\rm B\kern-.05em{\sc i\kern-.025em b}\kern-.08em
    T\kern-.1667em\lower.7ex\hbox{E}\kern-.125emX}}
\usepackage{xcolor}
\usepackage{listings}
\definecolor{promptgray}{RGB}{40,40,40}      
\definecolor{promptbg}{RGB}{248,248,248}     

\usepackage[table]{xcolor}
\definecolor{easyc}{RGB}{198,239,206}
\definecolor{mediumc}{RGB}{255,229,153}
\definecolor{hardc}{RGB}{244,204,204}

\definecolor{badc}{RGB}{244,204,204}
\definecolor{midbadc}{RGB}{252,229,205}
\definecolor{midgoodc}{RGB}{255,242,204}
\definecolor{goodc}{RGB}{217,234,211}

\lstdefinestyle{promptstyle}{
  basicstyle=\ttfamily\footnotesize,   
  columns=fullflexible,
  keepspaces=true,
  breaklines=true,
  showstringspaces=false,
  frame=single,
  rulecolor=\color{promptgray},
  backgroundcolor=\color{yellow!10},
  xleftmargin=0.25em,
  xrightmargin=0.25em,
  aboveskip=0.25em,
  belowskip=0.25em,
  commentstyle=\color{promptgray},
  keywordstyle=\color{promptgray},
  stringstyle=\color{promptgray},
  escapeinside={(*@}{@*)} 
}

\usetikzlibrary{fit}

\newcommand{\baditem}[1]{%
  \tikz[baseline=(X.base)]\node[
    draw=red,
    rounded corners=1pt,
    line width=0.4pt,
    inner xsep=1pt,
    inner ysep=0.2pt
  ] (X) {\ttfamily\fontsize{6}{6}\selectfont #1};%
}
\usepackage[normalem]{ulem}
\newcommand{\ignore}[1]{}

\definecolor{gray1}{gray}{0.7}

\newcommand\blfootnote[1]{%
\begingroup
\setlength{\skip\footins}{1pt}%
\renewcommand\thefootnote{}\footnote{\vspace{-0.4em}#1\vspace{-0.6em}}%
\addtocounter{footnote}{-1}%
\endgroup
}

\newcounter{promptsub}

\begin{document}



\title{{NEMESIS:} \underline{NE}tlist-Driven \underline{M}odeling and \underline{E}quation \underline{S}ynthesis with \underline{I}nversion-Aware \underline{S}PICE Anchoring}
\author{
\IEEEauthorblockN{Subhadip Ghosh, Ramesh Harjani, and Sachin S. Sapatnekar}
\IEEEauthorblockA{
\text{Department of Electrical and Computer Engineering, University of Minnesota, Minneapolis, MN, USA} 
}
}

\maketitle

\begin{abstract}
This work presents \textsc{NEMESIS}, a multimodal framework for operational transconductance amplifier (OTA) design using large language models (LLMs). \textsc{NEMESIS} strikes a balance between fast, approximate analytical models vs. accurate, computationally expensive SPICE evaluations. Given an OTA netlist and schematic\ignore{ for an OTA}, \textsc{NEMESIS} first identifies circuit primitives and then generates progressively more accurate performance equations. The framework begins with equations retrieved from the prior invocations of \textsc{NEMESIS} to structurally similar OTAs, if available; otherwise, it uses the LLM to derive the initial equations directly from the circuit input. These equations are iteratively refined via a SPICE-based repair loop. In a commercial 65nm PDK, NEMESIS is demonstrated on five OTA topologies, producing SPICE-verified equations across biasing ranges with $<7\%$ average relative error and a post-convergence evaluation speedup of $\approx4622\times$ over full SPICE-based evaluation. \blfootnote{This work was supported in part by the NSF under award 2212345.}
\end{abstract}

\section{Introduction}
\label{sec:introduction}

\noindent
Operational transconductance amplifiers (OTAs) are widely used in sensing, amplification, and data conversion, yet their design remains manual and iterative. Improving one performance target often perturbs others, and the performance bottleneck that prevents the design from meeting all specifications can shift with circuit topology, bias, and target requirements. This coupling motivates automated sizing and design-space exploration (DSE), which must evaluate many design candidates and bias points. Since each candidate is assessed across multiple metrics, repeated SPICE-based evaluation becomes expensive throughout the optimization loop. The performance model used in this loop, therefore, directly determines both runtime and design insight. Fast, explicit models can reduce this cost by linking device parameters to circuit-level performance, but generating such models remains difficult due to topology-specific dependencies and bias/small-signal effects.

Conventional approaches typically use one of two model classes for circuit-performance evaluation. The first uses explicit analytical equations derived from circuit structure.  
These models are fast and interpretable, but they are handcrafted, topology-specific, and often inaccurate in modern processes.
They include equation-driven models used in~\cite{abel_22_2, graeb_01, graeb_08, harjani_89,koh_90, Gebru26}, symbolic models invoked in~\cite{gielen_90}, and posynomial models used in geometric programming~\cite{daems_03,mandal_01,hershenson_98}. 
The second class directly uses SPICE for performance evaluation.
These evaluations are accurate and avoid analytical simplifications, but are computationally expensive and provide limited circuit-level insight.
This class has been used in optimizers based on Bayesian optimization~\cite{lyu_18,lyu_18_2,touloupas_21,gu_2024}, machine learning (ML)~\cite{budak_22, ghosh_2025}, and reinforcement learning~\cite{settaluri_20,wang_2020,budak_21,choi_23}. We aim to bridge these model classes with fast, interpretable equations that are SPICE-verified under a prescribed error threshold.

Beyond facilitating automated optimization,
a second motivation for our work is that better models could also aid manual DSE. 
For manual use, a model should be interpretable, with terms traceable to specific devices, nodes, and small-signal contributions rather than being hidden inside a black-box predictor.
Improved models link predicted behavior to circuit structure and aid both manual \ignore{DSE} and automated optimization.

\begin{figure*}[t]
    \centering
    \includegraphics[width=0.8\textwidth]{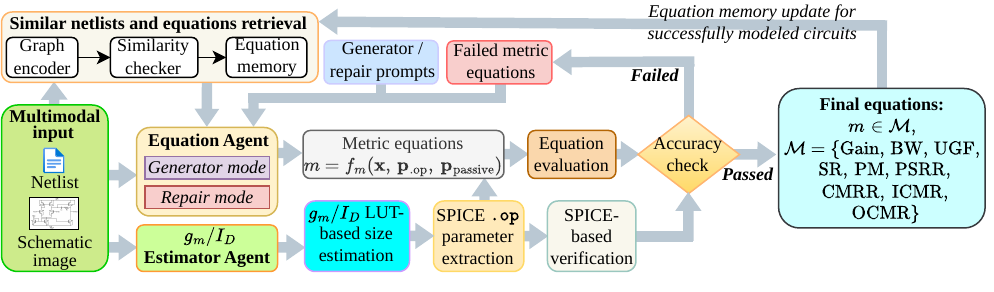}
    \vspace{-4mm}
    \caption{Overview of the proposed \textsc{NEMESIS} framework for automated OTA performance modeling. 
    }
    \label{fig:full_flow}
    \vspace{-5mm}
\end{figure*}

Our solution is based on using large language models (LLMs), which have been \ignore{deployed successfully}deployed in analog design workflows via copilots~\cite{lai_25}, search-oriented agents~\cite{yin_24, kochar_25}, and multi-agent assistants~\cite{liu_24, chen_24, shen_26}. 
LLMs leverage broad engineering knowledge to flexibly infer relationships among device parameters, internal nodes, and performance metrics, much like a human designer,
\ignore{These successes indicate} 
suggesting that they could be promising for accurate equation-based modeling. However, employing them requires an intermediate circuit representation that links the circuit description to the target metrics before the final equations are obtained, namely, a primitive-block decomposition of the circuit.
Primitive recognition has been studied through library-based~\cite{graeb_01, abel_22_2}, graph-learning~\cite{kunal_20}, and LLM methods~\cite{pham_25}. 
In\ignore{\textsc{NEMESIS}} this work, however, primitive recognition is not a standalone preprocessing step; it is an intermediate reasoning step within the same LLM pipeline that generates the performance equations.
This decomposition helps the model organize the circuit into meaningful primitive blocks before relating them to target metrics.

Our goal is to generate device-level, 
interpretable equations that are fast to evaluate, and accurate enough to replace repeated SPICE invocations during design optimization.
To address this problem, we present \textsc{NEMESIS}, a multimodal LLM framework for OTA performance modeling that bridges interpretable equations and closely aligns with SPICE. 
Leveraging primitive identification as an intermediate representation, we generate circuit-specific metric equations and employ retrieval-augmented generation (RAG)~\cite{lewis_20} over previously modeled, structurally similar OTAs to mitigate hallucinations.
In parallel, \textsc{NEMESIS} uses an estimator to assign 
each primitive a target transconductance efficiency ($g_m/I_D$) value~\cite{jespers_17,silviera_96} and an application-specific bias range intended to cover speed, balanced, or efficiency-oriented designer intent. 
These $g_m/I_D$ values guide OTA biasing and sizing by mapping to physically plausible device dimensions using technology-characterized lookup tables (LUTs).
Equations are first evaluated against SPICE and corrected, if necessary, at a central $g_m/I_D$ bias point in the range, and then checked over the assigned range.

Unlike
prior methods \ignore{that rely on external task-specific references at}\ignore{that use task-specific design references as context}that mine information from books, papers, manuals, etc.~\cite{lai_25, yin_24, kochar_25, liu_24, chen_24, shen_26}, 
\textsc{NEMESIS} \ignore{directly}uses a pretrained multimodal LLM--OpenAI GPT-5.2~\cite{gpt52}, \ignore{in our implementation}rather than starting with a foundation model and fine-tuning it on circuit-specific data.  This choice is made because such advanced LLMs are already capable of performing the tasks required by our flow without fine-tuning, and they continue to improve at a very rapid pace. 
A single model is shown to jointly reason across netlist text, schematic images, and metric contexts. 
LLM-generated equations are not accepted by default; \textsc{NEMESIS} uses structured prompts, equation-memory retrieval, and SPICE-guided repair, retaining equations only after circuit-level verification.
Our key contributions are as follows:

\begin{itemize}
  \item We present a multimodal framework for OTA equation generation from netlists and schematic images.
  \item We introduce a SPICE-guided loop for iterative equation verification and refinement to improve accuracy.
  \item We use the well-known relationship~\cite{jespers_17} between the region of operation of a device and the range of $g_m/I_D$ values, and ensure model accuracy over this range.
  \item We employ RAG with an equation memory to improve initialization, speed up repair, and limit hallucinations.
  \item We validate the framework on five OTA topologies that require progressively more complex performance models.
\end{itemize}
The remainder of this paper is organized as follows. Section~\ref{sec:framework} presents the \textsc{NEMESIS} framework, Section~\ref{sec:results} reports experimental results, and Section~\ref{sec:conclusion} concludes the paper.

\section{Proposed Framework: \textsc{NEMESIS}}
\label{sec:framework}
\noindent
Figure 1 shows the NEMESIS workflow, which derives a SPICE-verified equation-based OTA model from a SPICE netlist and schematic image. The generated equations are checked for accuracy and retained only if they satisfy an error threshold. We use the three-current-mirror OTA (OTA-2 in Table II) as a running example to demonstrate the framework.

The workflow proceeds in stages. First, an external equation memory retrieves structurally similar circuits with previously verified equations, if available, providing a grounded starting point for equation generation and reducing the likelihood of invalid expressions. 
Next, an \textit{Equation Agent} uses the netlist, schematic image, and retrieved context to generate initial equations for the requested metrics. In parallel, a companion \textit{$g_m/I_D$ Estimator Agent} assigns primitive-level admissible ranges for the $g_m/I_D$ values. 
Conventionally, the $g_m/I_D$ method is used for circuit optimization, but we use these values to map admissible values to concrete transistor dimensions via standard precharacterized LUTs. This process instantiates physically meaningful, technology-consistent operating points for SPICE evaluation of the accuracy of our models.

The primary SPICE-based repair loop uses the selected $g_m/I_D$ value for this instantiation. 
Once the equations match SPICE at this initial point, they are evaluated across the broader $g_m/I_D$ range; any significant localized errors trigger a secondary repair loop at the midpoint of the affected sub-range.
The framework runs OTA performance-estimation testbenches based on DC, AC, and transient analysis (Section~\ref{sec:spice_verify}) to obtain reference metric values, while also extracting the\ignore{operating-point and small-signal} parameters required for equation evaluation.
As compact approximations of circuit behavior, the generated equations may \ignore{not fully capture}omit the parasitic terms, higher-order effects, or bias-dependent interactions. Consequently, when predictions deviate from SPICE references beyond prescribed tolerances, \textsc{NEMESIS} invokes the \textit{Equation Agent} in \textit{repair mode}, utilizing the observed discrepancy to selectively revise only the failed metric equations. Together, these agents organize the workflow into three controlled operations: structured equation generation, primitive-level $g_m/I_D$ assignment for SPICE instantiation, and verification-driven equation repair.

\subsection{Graph-based equation memory retrieval}
\label{sec:rag}

\noindent
Before invoking the Equation Agent, \textsc{NEMESIS} queries an external equation memory if relevant, verified examples are already available from prior NEMESIS runs on other OTA topologies. This memory is populated progressively:
models for early, simple OTA topologies are generated without retrieval, and their SPICE-verified equations are stored to assist later, more complex designs.
For query circuits structurally similar to an earlier verified design, the retrieved context provides a grounded starting point for generation.
In this way, retrieval is used to improve initialization and reduce subsequent repair effort for more complex OTA topologies, building upon results from simpler topologies.

Each prior netlist is stored as
a bipartite labeled graph $G=(V_N \cup V_D, E)$, where $V_N$ denotes net nodes and $V_D$ denotes device nodes, connected by edges $E$ representing the device terminals. This representation removes dependency on instance names and sizes, preserves external pins as anchor nets to enforce connectivity-driven matching. 
For each graph, \textsc{NEMESIS} generates a compact structural signature consisting of a motif vector (capturing coarse block-level composition, e.g., shared-gate mirrors and device counts) and a two-iteration Weisfeiler-Lehman (WL)~\cite{wl_11} refinement hash (capturing local connectivity).
These features retrieve prior circuits with similar gain paths, loading, and biasing. Denoting the Jaccard similarity~\cite{jaccard_1901} of the WL hashes~\cite{wl_11} by $s_{\mathrm{WL}}$, and the cosine similarity of the motif vectors by $s_{\mathrm{motif}}$, the similarity between a query netlist $q$ and library entry $i$ is scored as:
\[
s(q,i)=\lambda s_{\mathrm{WL}}(q,i) + (1-\lambda)s_{\mathrm{motif}}(q,i),
\]
\ignore{where $s_{\mathrm{WL}}$ is the Jaccard similarity~\cite{jaccard_1901} of the WL hashes~\cite{wl_11}, $s_{\mathrm{motif}}$ is the cosine similarity of the motif vectors, and $\lambda=0.6$ balances topological and motif-level agreement.} where $\lambda$ balances topological and motif-level agreement; we choose $\lambda=0.6$. The most similar examples are then packed into the prompt, providing a bounded context of structural summaries and locked equations. Following SPICE verification, newly accepted equations are indexed via the same pipeline: thus, \ignore{ensuring} the database grows \ignore{exclusively} through verified updates.

\begin{figure}[t]
    \centering
    \begin{minipage}[t]{0.35\columnwidth}
        \vspace{0pt}
        \centering
        \includegraphics[width=\linewidth]{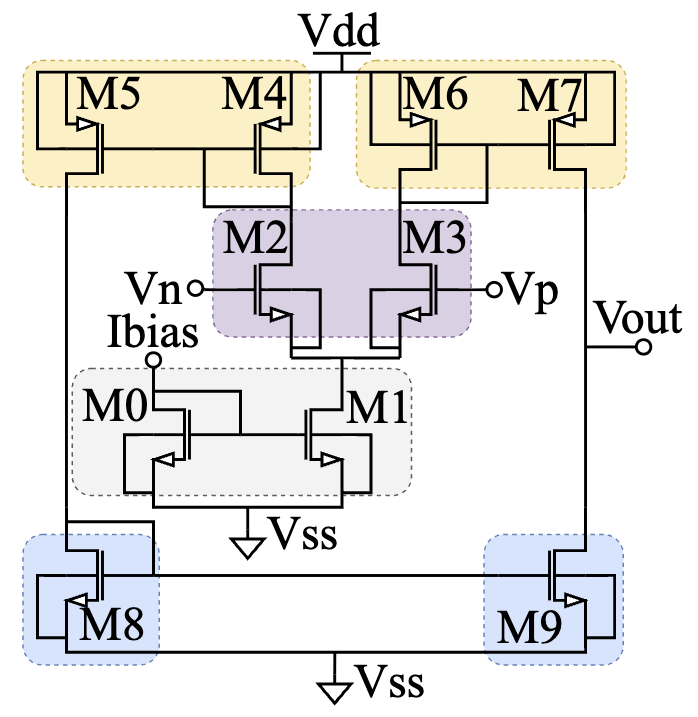}
    \end{minipage}
    \begin{minipage}[t]{0.6\columnwidth}
        \vspace{0pt}

\begin{lstlisting}[style=promptstyle,basicstyle=\ttfamily\fontsize{6}{6}\selectfont,breaklines=true,columns=fullflexible,  stringstyle=\color{promptgray}, escapeinside={(*@}{@*)},backgroundcolor=\color{yellow!10}]
(*@\textbf{Sample gm/Id estimator output for OTA-2}@*)
{(*@\textbf{"GMID\_TARGETS"}@*): {
 "DP_M2_M3":{"gmid":16,"range":[14,18]},
 "CM1_M4_M5":{"gmid":12,"range":[10,14]},
 "CM2_M6_M7":{"gmid":12,"range":[10,14]},
 "CM3_M8_M9":{"gmid":12,"range":[10,14]},
 "CM4_M0_M1":{"gmid":11,"range":[9,13]}},
 (*@\textbf{"REGION\_CONSTRAINTS"}@*): {
 "DP_M2_M3":"moderate (mod.) inversion",
 "CM1_M4_M5":"moderate inversion",
 "CM2_M6_M7":"moderate inversion",
 "CM3_M8_M9":"moderate inversion",
 "CM4_M0_M1":"mod.-to-strong inversion"}}
\end{lstlisting}
    \end{minipage}
    \caption{OTA-2 schematic and sample $g_m/I_D$-estimator output.}
    \label{fig:ota2_gmid_sidebyside}
    \vspace{-5mm}
\end{figure}

\subsection{LLM agents and prompt engineering}
\label{sec:inputs_prompts}

\noindent
The LLM agents are invoked through structured prompts. \ignore{, abbreviated here for space. For brevity, only abbreviated prompt templates are shown in this section.} 
Prompts combine two user inputs: (i)~a
circuit description, as a SPICE netlist and a schematic image, and (ii)~a target metric set $\mathcal{M}$. 
While the netlist establishes instance-level connectivity, the schematic offers complementary visual context to resolve port roles, mirrored branches, and otherwise ambiguous substructures~\cite{liu_25}.
Additionally, the Equation Agent receives retrieved memory examples as a bounded context. All agent responses are constrained to a strict JSON format so they can be parsed, checked, and immediately utilized. 

\begin{figure}[t]
\centering
\scriptsize

\begin{minipage}{\columnwidth}
\refstepcounter{promptsub}\label{fig:prompt_templates_gen}
\textbf{(a) \emph{Equation Agent:} generation mode}
\vspace{0.5mm}
\begin{lstlisting}[style=promptstyle,basicstyle=\ttfamily\fontsize{6.5}{5.5}\selectfont]
(*@\textbf{ROLE:}@*) Act as an expert analog IC designer ...
(*@\textbf{TASK:}@*) Analyze the netlist/schematic and output the
requested items strictly as a JSON object...
(*@\textbf{CRITICAL PRIORITY CHECK CONSTRAINT:}@*)
- If similar netlists are found in the database,
  use them to model the new netlist...
- Else, derive equations from connectivity...
(*@\textbf{STRICT PARAMETER USAGE:}@*)
- Use only device instances present in the netlist.
- Use only LUT parameter names as listed: gm, gds,
  gmb, gmid, cgs, cgd, cdd, cds, cbd, ... .
  Do not invent new symbols or hidden nodes.
(*@\textbf{OUTPUT FORMAT (STRICT JSON):}@*)
{ "1_TOPOLOGY_SUMMARY": {...},
  "2_DEVICE_MAPPING": {...},
  "3_PERFORMANCE_METRIC_EQUATIONS": {...},
  "4_PYTHON_EXECUTABLE": {...}}
(*@\textbf{NETLIST TO ANALYZE:}@*)
.subckt DUT <PIN1> <PIN2> ...
... (raw SPICE netlist) ...
.ends DUT
\end{lstlisting}
\end{minipage}


\begin{minipage}{\columnwidth}
\textbf{(b) \emph{$g_m/I_D$ Estimator Agent}}
\vspace{0.5mm}
\begin{lstlisting}[style=promptstyle,basicstyle=\ttfamily\fontsize{6.5}{5.5}\selectfont]
(*@\textbf{ROLE:}@*) Choose primitive-level gm/ID targets for...
(*@\textbf{INPUT:}@*) {netlist + schematic image + primitives...}
(*@\textbf{OUTPUT FORMAT (STRICT JSON):}@*) {T: group -> (gm/ID 
target, range), region constraints}...
\end{lstlisting}
\end{minipage}


\begin{minipage}{\columnwidth}
\textbf{(c) \emph{Equation Agent}: repair mode}
\vspace{0.5mm}
\begin{lstlisting}[style=promptstyle,basicstyle=\ttfamily\fontsize{6.5}{5.5}\selectfont]
(*@\textbf{ROLE:}@*) Verification-driven equation editor.
(*@\textbf{INPUT:}@*) {failed_metric, current_eq, error_summary,
bindings, locked_PASS_eqs}.
(*@\textbf{OUTPUT (STRICT JSON):}@*){patched_eq_for_failed_metric,
missing_term, brief reason}.
(*@\textbf{GUARDRAILS:}@*) DO NOT modify locked equations; use 
only LUT symbols; bounded edits.
\end{lstlisting}
\end{minipage}

\caption{Abbreviated prompt templates used in \textsc{NEMESIS}: (a) equation generation, (b) primitive-level $g_m/I_D$ estimation, and (c) verification-driven equation repair. For brevity, ellipses are used to denote omitted wording.}
\label{fig:prompt_templates}
\vspace{-5mm}
\end{figure}

The \textbf{first step} of \textsc{NEMESIS} is the \emph{generation mode} of the \emph{Equation Agent}. The agent is prompted, using the template shown in Figure~\ref{fig:prompt_templates}(a), to identify the circuit primitives, map devices, and generate executable equations for the target metric set $\mathcal{M}$. To prevent hallucination, the agent is restricted strictly to netlist instances. 
The output contains four machine-readable objects: 
(i)~a primitive decomposition topology summary, providing an intermediate structural abstraction to guide device-to-metric mapping and equation construction; 
(ii)~a device-level binding, mapping equation terms back to netlist instances, (iii)~an \ignore{executable} equation, for each $m \in \mathcal{M}$, of the form
\begin{equation}
m = f_m(\mathbf{x},\,\mathbf{p}_{\mathrm{.op}},\,\mathbf{p}_{\mathrm{passive}}),
\qquad \forall\, m\in\mathcal{M},
\end{equation}
and (iv) a Python-executable representation for downstream evaluation. 
Here, $\mathbf{x}$ captures the design-dependent variables, including device sizes and bias quantities. The remaining parameters are supplied later during evaluation: $\mathbf{p}_{\mathrm{.op}}$ denotes the 
operating-point parameters extracted from the SPICE \texttt{.op} analysis, and $\mathbf{p}_{\mathrm{passive}}$ are the passive component parameters.

The \textbf{second step} of \textsc{NEMESIS} is the \emph{$g_m/I_D$ estimator}. As shown in the prompt template (Figure~\ref{fig:prompt_templates}(b)), the agent processes the circuit description and the identified primitives to generate a mapping $\mathrm{T}$. This map assigns a selected $g_m/I_D$ value, an admissible range, and a coarse inversion-region label (weak, moderate, or strong) to each primitive group.
The mapping $\mathrm{T}$ \ignore{produced by this agent} drives the LUT-based sizing stage (Section~\ref{sec:gmid_sizing}) to instantiate transistor dimensions for SPICE evaluation. Beyond initial sizing, $\mathrm{T}$ anchors the primary repair loop, while the assigned ranges define the operating neighborhood for secondary verification. If significant errors occur within a sub-range, the midpoint of the affected sub-range is selected for another iteration of the SPICE-based repair. For OTA-2 (Figure~\ref{fig:ota2_gmid_sidebyside}), the differential pair receives the highest $g_m/I_D$, while the current-mirrors receive intermediate values, and the tail-bias branch is placed slightly closer to stronger inversion.

The \textbf{third step} of \textsc{NEMESIS} is the \emph{repair mode} of the \emph{Equation Agent}\ignore{. This mode is invoked only for metrics that fail SPICE verification, i.e., when the equation-predicted value differs from the corresponding SPICE reference by more than the error threshold.} invoked only for metrics that fail SPICE verification (i.e., those exceeding the error threshold).  The repair step is deliberately local: instead of regenerating the full equation set, the agent receives the failed metric, its current equation, the inaccuracy summary, and the locked bindings or equations that have already satisfied the error threshold, and is prompted (Figure~\ref{fig:prompt_templates}(c)) to propose a bounded update\ignore{only for the failing equations}.

\begin{table}[t]
\centering
\caption{SPICE verification suite and metric coverage used in \textsc{NEMESIS}.}
\label{tab:spice_tbs}
\vspace{-2mm}
\renewcommand{\arraystretch}{1.0}
\setlength{\tabcolsep}{1pt}
\scriptsize
\resizebox{\columnwidth}{!}{%
\begin{tabular}{|p{1.6cm}|p{5.4cm}|l|}
\hline
\textbf{Category} & \textbf{Metrics (full name with symbol)} & \textbf{Analysis} \\
\hline
DC sweep / Operating range &
Input and Output common-mode range (ICMR$_{min}$ - ICMR$_{max}$) and (OCMR$_{min}$ - OCMR$_{max}$) &
\texttt{.op} / \texttt{.dc} \\
\hline
AC / Stability &
DC gain (A$_{DC}$); 3-dB bandwidth (BW$_{3dB}$); unity-gain frequency (UGF); phase margin (PM); common-mode rejection ratio (CMRR); power-supply rejection ratio (PSRR$^{+}$/PSRR$^{-}$) &
\texttt{.ac} \\
\hline
Transient &
Positive/negative slew rate (SR$^{+}$/SR$^{-}$) &
\texttt{.tran} \\
\hline
\end{tabular}%
}
\vspace{-5mm}
\end{table}

\subsection{$g_m/I_D$ LUT-based sizing and region checks}
\label{sec:gmid_sizing}

\noindent
To enable SPICE-based evaluation, the primitive-level $g_m/I_D$ assignments are converted into physical transistor dimensions $W$ and $L$. 
For this step, \textsc{NEMESIS} uses the precharacterized lookup tables (LUTs) of the $g_m/I_D$ sizing methodology~\cite{jespers_17},  generated at a normalized width 
by sweeping channel length and terminal biases $(V_{GS},V_{DS},V_{SB})$. For each operating point, the tables store the quantities needed for sizing\ignore{and equation evaluation}, including $I_D$, $g_m$, $g_{ds}$, parasitic capacitances, 
$V_{DS\mathrm{sat}}$, and $V_{TH}$.

For a chosen $L$, NEMESIS selects LUT entries consistent with the assigned $g_m/I_D$ target and current level, then rescales the normalized width to obtain the physical $W$. Since $g_m/I_D$ primarily constrains the inversion level but not the drain--source headroom, the LUT lookup also considers the terminal-bias quantities stored during precharacterization, such as $V_{DS}$ and $V_{DS\mathrm{sat}}$. This avoids selecting entries that match $g_m/I_D$ and $I_D$ but correspond to a bias point clearly inconsistent with saturation. The check is therefore used to obtain a physically plausible first-cut sizing point for the instantiated OTA.

\subsection{SPICE-based verification}
\label{sec:spice_verify}

\noindent
Following LUT-based sizing, \textsc{NEMESIS} verifies the generated equations against SPICE for the same instantiated
circuit.  
For each sized netlist, it runs the HSPICE testbench suite~\cite{kunal_19,align_public_github} for the metrics in Table~\ref{tab:spice_tbs}: DC tests for input/output common-mode range, AC tests for gain, bandwidth, unity-gain frequency, phase margin, CMRR, and PSRR, and transient tests for slew rates.
A single \texttt{.op} analysis extracts operating-point/device parameters needed to evaluate the generated equations. This avoids a key limitation of handcrafted first-order equations, whose fixed-region assumptions and missing bias dependence can cause more than $20\%$ SPICE-relative error for several OTA metrics. Using \texttt{.op}-extracted small-/large-signal parameters from the actual operating point, \textsc{NEMESIS} captures bias-/region-dependent behavior while keeping equations explicit and comparing predictions to SPICE.\ignore{under the same sizes and bias.}

Verification is performed independently for each metric $m \in \mathcal{M}$ 
using a relative-error threshold $\epsilon$. 
Errors may arise from omitted parasitics, higher-order effects, or operating-point interactions captured by SPICE but absent from the current equation form.
A metric is marked \texttt{PASS} if the equation prediction matches SPICE within $\epsilon$ and \texttt{FAIL} otherwise; the threshold, therefore, serves as the acceptance criterion for whether the current equation is sufficiently accurate for the sized circuit. For each failed metric, \textsc{NEMESIS} records the equation output, SPICE reference, and mismatch, and sends them to the \emph{repair mode} of the \emph{Equation Agent}. Equations that pass are retained unchanged, so that subsequent updates are restricted to only the failed equations.
In the default flow, repair is performed at the circuit instantiation set by the selected $g_m/I_D$ values. After the equations satisfy the threshold there, they are checked at additional points in the assigned $g_m/I_D$ ranges; if a significant error appears, repair is repeated at the midpoint of the affected sub-range.

\section{Experimental Results}
\label{sec:results}
\noindent

\begin{table}[t]
\centering
\caption{OTA topologies used to evaluate \textsc{NEMESIS}.}
\label{tab:topo_summary}
\vspace{-2mm}
\scriptsize
\setlength{\tabcolsep}{1pt}
\renewcommand{\arraystretch}{1.05}
\resizebox{\columnwidth}{!}{%
\begin{tabular}{|c|p{2.2cm}|p{5.8cm}|}
\hline
\textbf{ID} & \textbf{Topology} & \textbf{Primitive blocks (count)} \\
\hline \hline
\cellcolor{easyc}\textbf{OTA-1} & 5-transistor OTA &
DP (1), CM (1), biasing CM (1) \\
\hline
\cellcolor{mediumc}\textbf{OTA-2} & 3-current-mirror OTA &
DP (1), CM (3), biasing CM (1) \\
\hline
\cellcolor{mediumc}\textbf{OTA-3} & 2-stage OTA &
DP (1), CM (1), CS (1), RCCN (1), biasing CM (2) \\
\hline
\cellcolor{hardc}\textbf{OTA-4} & LV cascode OTA &
DP (1), CM (1), LV cascode (2), biasing CM (1) \\
\hline
\cellcolor{hardc}\textbf{OTA-5} & Telescopic OTA &
DP (1), cascode CM (2), stacked CD (1), biasing CM (1) \\
\hline
\end{tabular}%
}
\vspace{1mm}

\parbox{\columnwidth}{\scriptsize
\textbf{Difficulty:}
\colorbox{easyc}{\strut Easy},
\colorbox{mediumc}{\strut Medium},
\colorbox{hardc}{\strut Hard};
DP: differential pair;
CM: current mirror;
CS: common-source stage;
RCCN: resistor-capacitor compensation network;
LV: low-voltage;
CD: cascode devices.
}
\vspace{-3mm}
\end{table}

\subsection{Experimental setup}
\label{sec:results_setup}
\noindent
\textsc{NEMESIS} is evaluated on five OTA topologies listed in Table~\ref{tab:topo_summary}. We group the OTAs into ``easy'', ``medium'', and ``hard'' classes based on topology complexity and the expected difficulty of deriving compact, accurate equations across the target metrics. All circuit simulations are performed in HSPICE using a commercial 65nm PDK with $V_{DD}=1\,\mathrm{V}$\ignore{and $V_{SS}=0\,\mathrm{V}$}.
All LLM-based stages use OpenAI GPT-5.2~\cite{gpt52} with its temperature parameter set to 0.2, and LLM-dependent results are averaged over five independent runs per OTA.

For OTA-1, OTA-2, and OTA-3, equation generation is performed without retrieval augmentation, and the resulting verified equations are used to seed the equation memory context for the harder OTAs, i.e., OTA-4 and OTA-5. For each target metric, convergence is declared when the relative error w.r.t. SPICE falls below $\epsilon$. We use $\epsilon=15\%$ as the stopping threshold for iterative repair; the average final error is well below this value, and the threshold governs only repair termination while final assessment remains SPICE-based. For every topology, we track the first converged iteration, the full repair trajectory, and the final converged error. The open-source \textsc{NEMESIS} repository, including prompts and scripts, is available at \url{https://github.com/UMN-EDA/NEMESIS}.

\begin{figure}[t]
\centering
\includegraphics[width=\columnwidth]{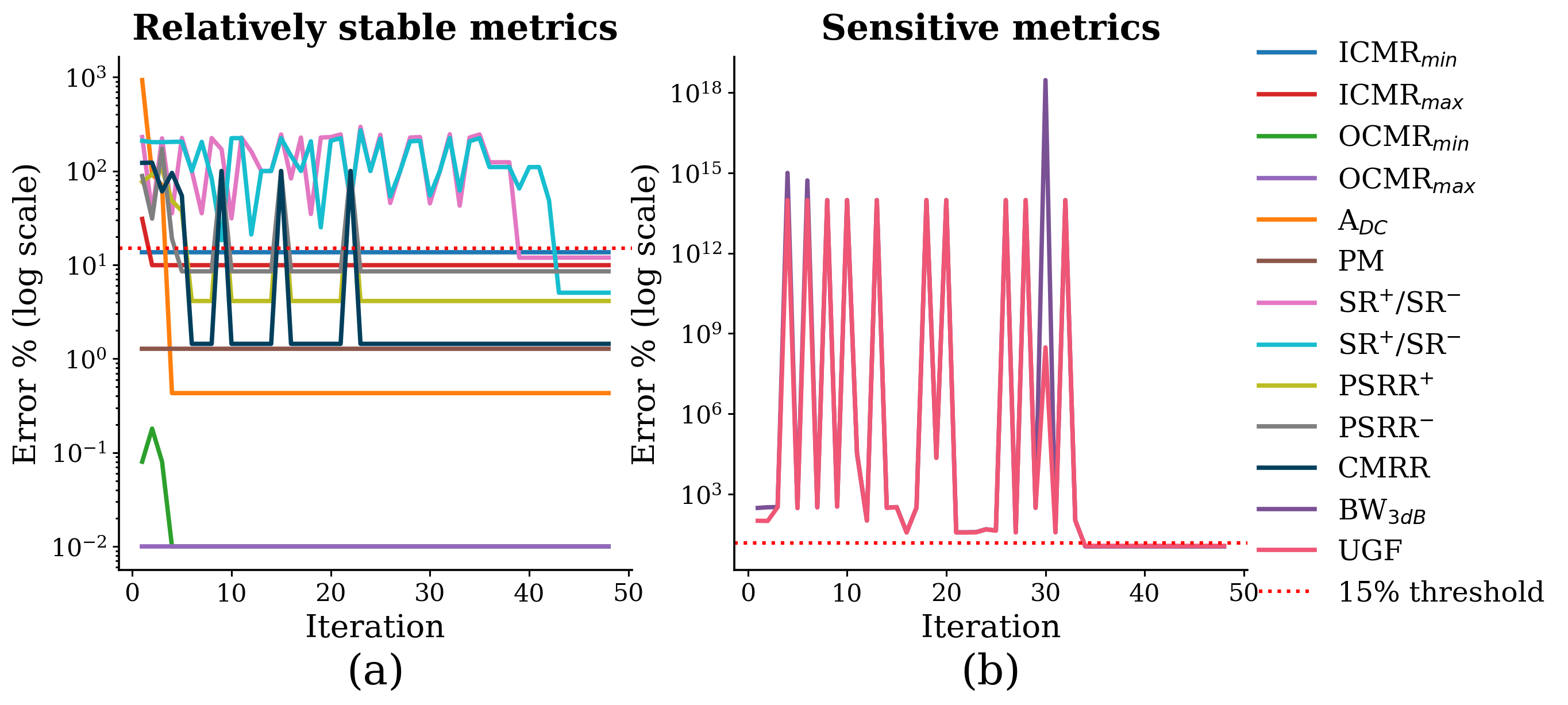}
\vspace{-8mm}
\caption{Error trajectories across equation repair iterations for OTA-2 for (a) relatively stable metrics and (b) sensitive metrics (UGF and BW$_{3dB}$).}

\label{fig:error_iter_rep}
\vspace{-6mm}
\end{figure}

\subsection{Convergence under iterative repair}
\label{sec:results_convergence}

\noindent
We examine the evolution of the generated equations under the repair loop by tracking their relative error. Figure~\ref{fig:error_iter_rep} shows these error trajectories for OTA-2, while Table~\ref{tab:conv_iter_err_all} reports the convergence iteration and final SPICE-relative error for all five OTAs. As shown in the figure, the repair loop reduces error across successive iterations, although the trajectories are not strictly monotone, since a bounded revision that improves one part of an equation can perturb other coupled terms before later iterations restore accuracy. 
Sensitive metrics like $BW_{3dB}$ and UGF show larger cross-iteration variation (Figure~\ref{fig:error_iter_rep}(b)) than the relatively stable metrics (Figure~\ref{fig:error_iter_rep}(a)). This aligns with their stronger dependence on pole locations and parasitic capacitances, whereas metrics governed by fewer parameters exhibit smoother repair trajectories.

Table~\ref{tab:conv_iter_err_all} shows that convergence is eventually reached for all five OTAs, with average final relative error below $7\%$ in every case. The required number of repair iterations, however, varies substantially. Considering first the
initial testcases before NEMESIS has prior OTA models that can be used for RAG, OTA-1 converges with the least repair effort, consistent with its simple
topology, while OTA-2 and OTA-3 require more iterations, indicating greater difficulty in equation generation. For example, OTA-3 is a two-stage OTA that extends the OTA-1 core with an additional common-source stage and therefore serves as a useful intermediate-complexity case; together, OTA-1 through OTA-3 provide a baseline for repair behavior without retrieval. Retrieval is then assessed in OTA-4 and OTA-5, which are initialized using verified equations retrieved from the 
equation memory.
Despite being harder topologies, both converge in fewer repair iterations than OTA-2 and OTA-3, consistent with retrieval providing a stronger initialization and reducing repair effort.

\begin{table}[t]
\centering
\caption{Number of iterations to convergence (Iter.) and final relative error (Err.) (\%) for each metric across OTA topologies.}
\vspace{-2mm}
\label{tab:conv_iter_err_all}
\vspace{1pt}
\scriptsize
\setlength{\tabcolsep}{1pt}
\renewcommand{\arraystretch}{1.0}
\resizebox{\columnwidth}{!}{%
\begin{tabular}{|l|r|r|r|r|r|r|r|r|r|r|}
\hline
\multirow{2}{*}{\textbf{Metric}} 
& \multicolumn{2}{c|}{\cellcolor{easyc}\textbf{OTA-1}} 
& \multicolumn{2}{c|}{\cellcolor{mediumc}\textbf{OTA-2}} 
& \multicolumn{2}{c|}{\cellcolor{mediumc}\textbf{OTA-3}} 
& \multicolumn{2}{c|}{\cellcolor{hardc}\textbf{OTA-4}} 
& \multicolumn{2}{c|}{\cellcolor{hardc}\textbf{OTA-5}} \\
\cline{2-11}
& \textbf{Iter.} & \textbf{\% Err.} 
& \textbf{Iter.} & \textbf{\% Err.} 
& \textbf{Iter.} & \textbf{\% Err.} 
& \textbf{Iter.} & \textbf{\% Err.} 
& \textbf{Iter.} & \textbf{\% Err.} \\
\hline \hline
A$_{DC}$      & 2  & 2.57\%  & 4  & 0.43\%   & 3  & 0.54\%   & 4  & 6.80\%   & 1  & 5.87\%   \\ \hline
PM            & 8 & 0.17\%   & 3  & 1.27\%   & 3  & 8.91\%   & 15 & 0.51\%   & 6  & 2.31\%   \\ \hline
SR$^{+}$      & 12 & 10.86\%  & 43 & 11.91\%  & 4  & 10.91\%  & 14 & 11.03\%  & 16 & 11.88\%  \\ \hline
SR$^{-}$      & 11 & 10.54\%  & 39 & 8.28\%   & 1  & 4.14\%   & 3  & 9.08\%   & 14 & 9.45\%   \\ \hline
CMRR          & 2  & 9.61\%   & 2  & 1.44\%   & 13 & 1.54\%   & 10 & 6.83\%   & 17 & 14.22\%  \\ \hline
PSRR$^{+}$    & 7  & 9.44\%   & 6  & 4.12\%   & 4  & 2.52\%   & 10 & 6.53\%   & 25 & 4.09\%   \\ \hline
PSRR$^{-}$    & 9  & 4.37\%   & 5  & 8.53\%   & 28 & 10.57\%  & 4  & 0.59\%   & 1  & 9.72\%   \\ \hline
ICMR$_{min}$  & 8  & 5.63\%   & 43 & 13.66\%  & 43 & 0.03\%   & 15 & 13.64\%  & 12 & 9.77\%   \\ \hline
ICMR$_{max}$  & 1  & 13.68\%  & 2  & 9.92\%   & 1  & 5.06\%   & 2  & 9.81\%   & 2  & 11.72\%  \\ \hline
OCMR$_{min}$  & 3  & 5.05\%   & 1  & 1.59\%   & 3  & 0.02\%   & 12 & 0.81\%   & 4  & 0.23\%   \\ \hline
OCMR$_{max}$  & 2  & 3.15\%   & 4  & 4.75\%   & 1  & 0.03\%   & 2  & 0.17\%   & 1  & 0.81\%   \\ \hline
BW$_{3dB}$    & 12 & 4.43\%   & 34 & 5.97\%   & 44 & 6.51\%   & 18 & 0.71\%   & 14 & 1.59\%   \\ \hline
UGF           & 12 & 5.52\%  & 34 & 5.98\%   & 39 & 7.49\%   & 11 & 4.74\%   & 12 & 5.99\%   \\ \hline
\hline
\makecell{\textbf{Max Iter. /} \\ \textbf{Avg. Err. (\%)}} 
& \textbf{12} & \textbf{6.54\%} 
& \textbf{43} & \textbf{5.98\%} 
& \textbf{44} & \textbf{4.28\%} 
& \textbf{18} & \textbf{5.48\%} 
& \textbf{25} & \textbf{6.74\%} \\
\hline
\end{tabular}}
\vspace{-3mm}
\end{table}

\subsection{Accuracy across application-specific $g_m/I_D$ range}
\label{sec:results_generalization}

\noindent
Next, we evaluate whether the converged equations remain accurate over the assigned $g_m/I_D$ bias range of each OTA, which spans the intended bias space of the topology from higher-speed through balanced to higher-efficiency operation. Starting from the equations that converge at the selected $g_m/I_D$ value, we realize the nominal point together with additional representative points spanning the full primitive-level $g_m/I_D$ ranges, including boundary and interior cases, using the LUT-based sizing flow of Section~\ref{sec:gmid_sizing}. The equations are then reevaluated at each realized operating point using updated \texttt{.op}-extracted parameters. In the evaluated testcases, no additional repair was required within the assigned ranges, since the equations that converged at the selected $g_m/I_D$ value remained accurate throughout the evaluated bias space.

\begin{figure}[t]
\centering
    \includegraphics[width=\columnwidth]{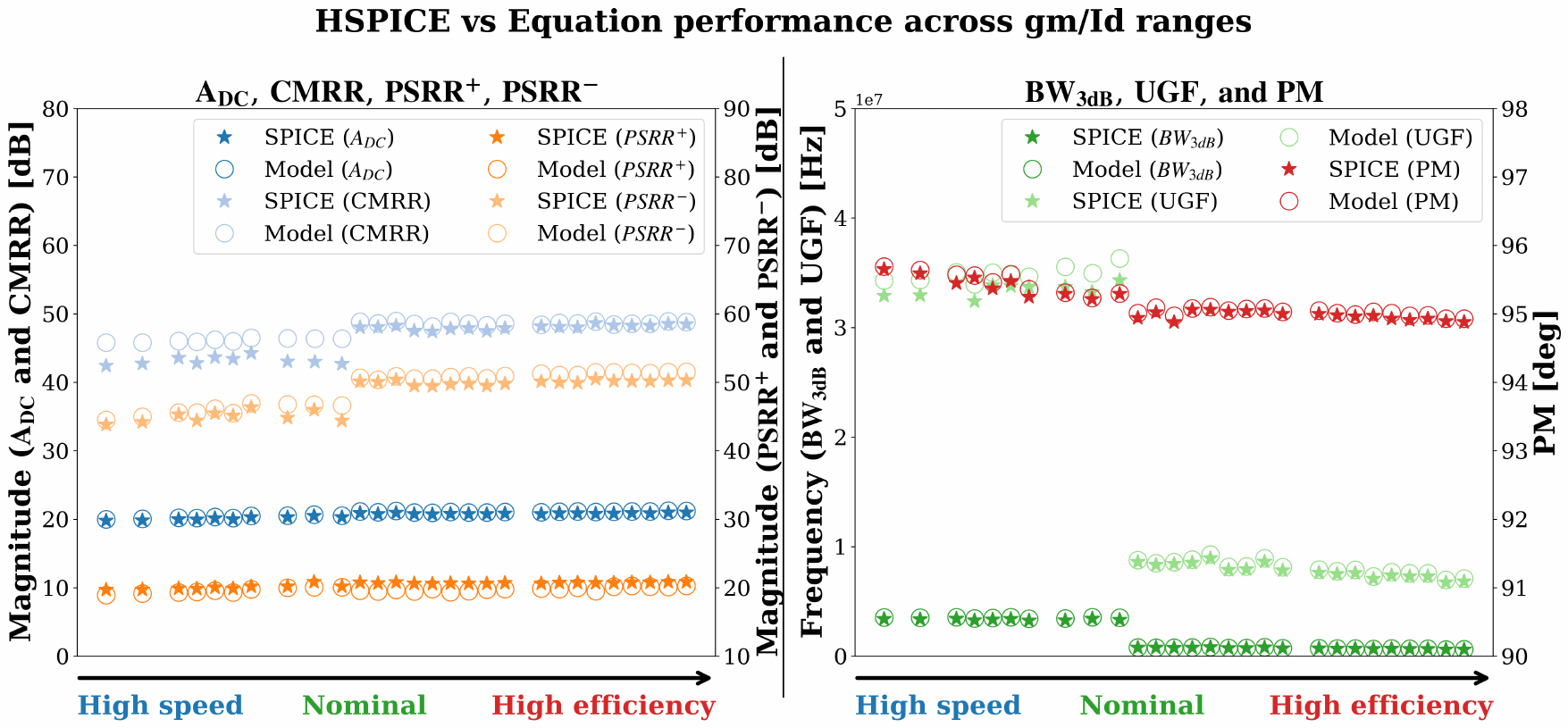}
\caption{Accuracy of the converged equations for OTA-2 across representative $g_m/I_D$ sweep points spanning different inversion regions.}
\label{fig:intent_region_acc}
\vspace{-6mm}
\end{figure}

Figure~\ref{fig:intent_region_acc} shows representative OTA-2 metrics, for which the equation-based predictions remain close to SPICE across the evaluated $g_m/I_D$ range. The full results for all OTAs are summarized in Table~\ref{tab:error_stats_all}. Across the evaluated $g_m/I_D$ ranges, the average mean relative error remains below $7\%$ for all five OTAs, with many metrics showing substantially lower error. Overall, the converged equations remain usable across the application-specific $g_m/I_D$ ranges.

\begin{table}[t]
\centering
\caption{Mean ($\mu$) and standard deviation ($\sigma$) of \% error of the equations across the application-specific $g_m/I_D$ range.}
\vspace{-2mm}
\label{tab:error_stats_all}
\vspace{1pt}
\scriptsize
\setlength{\tabcolsep}{2pt}
\renewcommand{\arraystretch}{1.0}
\resizebox{\columnwidth}{!}{%
\begin{tabular}{|l|r|r|r|r|r|r|r|r|r|r|}
\hline
\multirow{2}{*}{\textbf{Metric}} 
& \multicolumn{2}{c|}{\cellcolor{easyc}\textbf{OTA-1}} 
& \multicolumn{2}{c|}{\cellcolor{mediumc}\textbf{OTA-2}} 
& \multicolumn{2}{c|}{\cellcolor{mediumc}\textbf{OTA-3}} 
& \multicolumn{2}{c|}{\cellcolor{hardc}\textbf{OTA-4}} 
& \multicolumn{2}{c|}{\cellcolor{hardc}\textbf{OTA-5}} \\
\cline{2-11}
& \multicolumn{1}{c|}{\textbf{$\mu$}}
& \multicolumn{1}{c|}{\textbf{$\sigma$}}
& \multicolumn{1}{c|}{\textbf{$\mu$}}
& \multicolumn{1}{c|}{\textbf{$\sigma$}}
& \multicolumn{1}{c|}{\textbf{$\mu$}}
& \multicolumn{1}{c|}{\textbf{$\sigma$}}
& \multicolumn{1}{c|}{\textbf{$\mu$}}
& \multicolumn{1}{c|}{\textbf{$\sigma$}}
& \multicolumn{1}{c|}{\textbf{$\mu$}}
& \multicolumn{1}{c|}{\textbf{$\sigma$}} \\
\hline \hline
A$_{DC}$      & 1.15   & 0.08  & 0.54  & 0.34  & 0.52  & 0.05  & 2.51 & 1.45 & 1.89 & 0.76 \\ \hline
PM            & 0.04   & 0.02  & 1.01  & 0.22  & 9.93  & 0.95  & 3.55 & 0.84 & 2.41 & 1.44 \\ \hline
SR$^{+}$      & 4.64   & 3.49  & 9.89  & 5.21  & 13.66 & 2.43  & 7.89 & 4.11 & 11.12 & 2.69 \\ \hline
SR$^{-}$      & 9.04   & 1.97  & 9.65  & 4.39  & 3.24  & 0.78  & 8.88 & 2.69 & 9.8 & 1.01 \\ \hline
CMRR          & 3.61   & 3.88  & 3.89  & 1.30  & 1.73  & 1.53  & 2.73 & 0.58 & 3.17& 0.95 \\ \hline
PSRR$^{+}$    & 4.14   & 1.00  & 1.86  & 1.98  & 1.19  & 0.39  &3.41 & 1.66 & 5.2 & 2.04 \\ \hline
PSRR$^{-}$    & 1.92   & 0.59  & 4.07  & 2.00  & 7.75 & 1.24  &3.55 &2.03 & 4.59 & 3.61 \\ \hline
ICMR$_{min}$  & 3.67   & 4.48  & 9.97  & 5.09  & 9.04  & 2.31  & 7.54 & 6.12 & 5.17 & 4.68 \\ \hline
ICMR$_{max}$  & 0.32   & 1.42  & 3.36  & 1.67  & 5.69  & 0.86  & 1.58 & 0.81 & 2.66 & 3.22 \\ \hline
OCMR$_{min}$  & 10.90  & 3.07  & 7.65  & 1.77  & 4.56    & 1.98    & 5.10 & 0.47 & 4.59 & 2.41 \\ \hline
OCMR$_{max}$  & 1.92   & 6.59  & 3.14  & 4.27  & 3.15  & 1.10  & 1.08 & 0.39 & 1.77 & 0.98 \\ \hline
BW$_{3dB}$    & 3.57   & 1.27  & 4.86  & 3.59  & 8.44  & 0.50  & 9.84 & 2.91 & 8.19 & 4.36 \\ \hline
UGF           & 3.85   & 0.30  & 4.98  & 1.58  & 12.73 & 2.74 & 11.79 & 4.51 & 5.88 &6.59 \\
\hline
\end{tabular}}
\vspace{-3mm}
\end{table}

\begin{table}[t]
\centering
\caption{Runtime comparison: Full HSPICE versus equation-based evaluation with \texttt{.op} and equation breakdown. }
\vspace{-2mm}
\label{tab:runtime}
\vspace{1pt}
\scriptsize
\setlength{\tabcolsep}{2pt}
\renewcommand{\arraystretch}{1.08}
\resizebox{\columnwidth}{!}{%
\begin{tabular}{|c||c||c|c|c||c|}
\hline
\textbf{Topology}
& \makecell{\textbf{SPICE-based} \\ \textbf{evaluation}}
& \makecell{\textbf{\texttt{.op}} \\ \textbf{evaluation}}
& \makecell{\textbf{Equation} \\ \textbf{evaluation}}
& \makecell{\textbf{Total} \\}
& \makecell{\textbf{Speedup}} \\
\hline \hline
OTA-1 & 6.74s & 1.08ms & 0.22ms & 1.30ms & 5184\(\times\) \\ \hline
OTA-2 & 6.89s & 1.34ms & 0.29ms & 1.63ms & 4226\(\times\) \\ \hline
OTA-3 & 6.99s & 1.16ms & 0.28ms & 1.44ms & 4854\(\times\) \\ \hline
OTA-4 & 7.05s & 1.28ms & 0.31ms & 1.59ms & 4434\(\times\) \\ \hline
OTA-5 & 7.11s & 1.31ms & 0.30ms & 1.61ms & 4416\(\times\) \\
\hline
\end{tabular}%
}
\vspace{-3mm}
\end{table}

\begin{figure}[t]
\centering
\scriptsize
\begin{tikzpicture}[
    >=Latex,
    node distance=2.2mm,
    eqbox/.style={
        draw,
        rounded corners=1pt,
        align=left,
        inner sep=2pt,
        text width=0.9\columnwidth,
        font=\fontsize{6}{6.4}\selectfont,
        execute at begin node={
            \setlength{\arraycolsep}{0pt}
            \mathsurround=0pt
            \thinmuskip=0mu
            \medmuskip=0mu
            \thickmuskip=0mu
        }
    },
    errbubble/.style={
        draw=red!80!black,
        fill=red!12,
        circle,
        minimum size=7.5mm,
        inner sep=0pt,
        font=\scriptsize\selectfont\bfseries,
        text=red!80!black
    }
]

\node[eqbox, fill=badc] (eq1) {%
$\begin{array}{@{}l@{}}
A_{DC}^{(1)}=20\log_{10}|((g_{m_{M3}}+g_{mb_{M3}})-(g_{m_{M2}}+g_{mb_{M2}}))/\\[1pt]
\quad\quad\quad\quad(g_{ds_{M7}}+g_{ds_{M9}})|
\end{array}$};

\node[errbubble, right=1mm of eq1.east] (e1) {929\%};

\node[eqbox, fill=midbadc, below=of eq1] (eq2) {%
$\begin{array}{@{}l@{}}
A_{DC}^{(2)}=20\log_{10}|((g_{m_{M2}}+g_{m_{M3}})/(g_{ds_{M2}}+g_{ds_{M3}}+g_{ds_{M4}}+g_{ds_{M5}}))\\[1pt]
\quad\quad\quad\quad\times(g_{m_{M7}}/(g_{ds_{M7}}+g_{ds_{M9}}))|
\end{array}$};

\node[errbubble, right=1mm of eq2.east] (e2) {92\%};

\node[eqbox, fill=midgoodc, below=of eq2] (eq3) {%
$\begin{array}{@{}l@{}}
A_{DC}^{(3)}=20\log_{10}|((g_{m_{M2}}+g_{m_{M3}})/(g_{ds_{M2}}+g_{ds_{M3}}+g_{ds_{M2}}+g_{ds_{M3}}\\[1pt]
\quad\quad\quad\quad+g_{ds_{M4}}+g_{ds_{M5}}+g_{ds_{M6}}))\times(g_{m_{M7}}/(g_{ds_{M7}}+g_{ds_{M9}}))|
\end{array}$};

\node[errbubble, right=1mm of eq3.east] (e3) {71\%};

\node[eqbox, fill=goodc, below=of eq3] (eq4) {%
$\begin{array}{@{}l@{}}
A_{DC}^{(4)}=20\log_{10}|((g_{m_{M2}}+g_{m_{M3}})/(g_{ds_{M2}}+g_{ds_{M3}}+g_{m_{M4}}+g_{ds_{M4}}\\[1pt]
\quad\quad\quad\quad+g_{ds_{M5}}+g_{m_{M6}}+g_{ds_{M6}}))\times(g_{m_{M7}}/(g_{ds_{M7}}+g_{ds_{M9}}))|
\end{array}$};

\node[errbubble, right=1mm of eq4.east] (e4) {0.43\%};

\draw[->, thick] (eq1.south) -- (eq2.north);
\draw[->, thick] (eq2.south) -- (eq3.north);
\draw[->, thick] (eq3.south) -- (eq4.north);

\end{tikzpicture}
\vspace{-6mm}
\caption{Evolution of the OTA-2 DC-gain equation across repair iterations 1 to 4. Red circles indicate the relative error at each iteration.}
\label{fig:eq_evolution_adc_ota2}
\vspace{-5mm}
\end{figure}

\begin{figure}[t]
    \centering
    \begin{minipage}[t]{0.48\columnwidth}
        \vspace{0pt}
        \centering
        \textbf{Netlist only}
        \vspace{-2mm}

\begin{tikzpicture}[baseline=(box.north)]
\node[inner sep=0, outer sep=0] (box) {%
\begin{minipage}[t]{\linewidth}
\begin{lstlisting}[style=promptstyle,basicstyle=\ttfamily\fontsize{6}{6}\selectfont,breaklines=true,columns=fullflexible,showstringspaces=false,escapeinside={(*@}{@*)}]
{(*@\texttt{\bfseries "diff\_pair"}@*):[M2, M3],
 (*@\texttt{\bfseries "active\_load"}@*):(*@\baditem{[M4, M5, M7]}@*),
 (*@\texttt{\bfseries "tail\_current"}@*):[M1],
 (*@\texttt{\bfseries "bias\_reference"}@*):[M0, (*@\baditem{M8]}@*),
 (*@\baditem{\texttt{\bfseries "second\_stage"}}@*):[M8, M9]}
\end{lstlisting}
\end{minipage}%
};
\begin{scope}[overlay]
\node[anchor=north east, xshift=1mm, yshift=-5mm] at (box.north east)
{\color{red}\huge\ding{55}};
\end{scope}
\end{tikzpicture}
    \end{minipage}\hspace{1mm}
    \begin{minipage}[t]{0.48\columnwidth}
        \vspace{0pt}
        \centering
        \textbf{Netlist + schematic image}
        \vspace{-2mm}

\begin{tikzpicture}[baseline=(box.north)]
\node[inner sep=0, outer sep=0] (box) {%
\begin{minipage}[t]{\linewidth}
\begin{lstlisting}[style=promptstyle,basicstyle=\ttfamily\fontsize{6}{6}\selectfont,breaklines=true,columns=fullflexible,showstringspaces=false]
{(*@\texttt{\bfseries "diff\_pair"}@*):[M2, M3],
 (*@\texttt{\bfseries "active\_load"}@*):[M4, M5, M6, M7],
 (*@\texttt{\bfseries "tail\_current"}@*):[M1],
 (*@\texttt{\bfseries "bias\_mirror"}@*):[M0, M1],
 (*@\texttt{\bfseries "output\_sink"}@*):[M8, M9]}
\end{lstlisting}
\end{minipage}%
};
\begin{scope}[overlay]
\node[anchor=north east, xshift=1mm, yshift=-5mm] at (box.north east)
{\color{green!60!black}\huge\ding{51}};
\end{scope}
\end{tikzpicture}
    \end{minipage}
    \caption{Primitive-device mappings predicted by the LLM for OTA-2 under netlist-only and netlist + schematic-image inputs.}
    \label{fig:ota2_primitive_ablation}
    \vspace{-3mm}
\end{figure}

\subsection{NEMESIS vs. SPICE-based evaluation runtime}
\label{sec:results_runtime}
\noindent
We compare post-convergence evaluation in \textsc{NEMESIS} against the full HSPICE verification flow, excluding the one-time model-construction stage. In the SPICE baseline, all target metrics are obtained from the metric-specific testbenches described in Section~\ref{sec:spice_verify}. After convergence, \textsc{NEMESIS} estimates the same metrics using one \texttt{.op} extraction, which takes only a few milliseconds, followed by direct evaluation of the converged equations.

As detailed in Table~\ref{tab:runtime}, the \textsc{NEMESIS}-generated OTA model yields a 4226$\times$--5184$\times$ speedup over the complete SPICE testbench flow by computing all metrics at a single realized operating point rather than repeating full DC/AC/TRAN sweeps. During the excluded model-construction stage, each LLM call typically took 30sec--2min, which is dominated by API latency and server traffic.

\subsection{Interpretability through equation evolution}
\label{sec:results_interpretability}
\noindent
In addition to improving accuracy, the \textsc{NEMESIS} repair loop preserves interpretability by expressing each repair in device-level small-signal terms.
Figure~\ref{fig:eq_evolution_adc_ota2} shows this behavior for the DC-gain equation of OTA-2 over four repair iterations.
In this example, 
Iteration~1 is an initial LLM-generated equation lacking reliable circuit interpretation, reflecting the incompleteness or hallucinated term selection typical of a first-pass symbolic guess before SPICE-grounded repair. 

From Iteration~2 onward, the equation connects directly to the circuit structure. For example, the $g_{ds_{M7}}+g_{ds_{M9}}$ term is physically meaningful because M7 and M9 connect to the output node, contributing to the small-signal conductance seen at $v_{out}$. Later iterations introduce additional conductance terms, clarifying that gain is affected by the conductances associated with the output node and internal nodes in the bias and mirror network. The inclusion of $g_{m_{M4}}$ and $g_{m_{M6}}$ in the final iteration is also circuit-consistent, as M4 and M6 are diode-connected PMOS devices whose small-signal transconductance contributes to the conductance at the corresponding internal nodes. Ultimately, interpretability stems from the repaired equation itself: as the expression is corrected, it becomes both more accurate with respect to SPICE and more useful for manual analysis, with terms remaining traceable to specific devices and node loading in the topology.

\subsection{Ablation and sensitivity studies}
\label{sec:ablation}
\noindent
We present three studies to assess key components of \textsc{NEMESIS}: multimodal circuit input, equation memory retrieval, and the convergence threshold $\epsilon$ in the repair loop.

\noindent
\textbf{Ablation on multimodal input:}
Figure~\ref{fig:ota2_primitive_ablation} compares primitive extraction for OTA-2 using netlist-only and netlist-plus-schematic inputs. With netlist-only input, the LLM omits M6 from the active-load group, misassigns M8 to the bias current mirror, and introduces a nonexistent \texttt{second\_stage} block. Adding the schematic image corrects these errors and aligns the primitive groups with the intended topology. Similar netlist-only errors were observed across all OTAs, indicating that multimodal input improves structural grounding and reduces topology hallucinations during primitive extraction.

\noindent
\textbf{Ablation on equation memory retrieval:}
To isolate the effect of retrieval, we repeated OTA-4 and OTA-5 without equation memory. The repair iterations increased from 18 to 61 for OTA-4 and from 25 to 57 for OTA-5. Retrieval primarily improves initialization and reduces repair effort.

\begin{table}[t]
\centering
\caption{Number of iterations (Iter.) and percentage error (Err.) w.r.t. SPICE for OTA-1 for each metric, for $\epsilon=15\%$ and $\epsilon=5\%$.}
\vspace{-2mm}
\label{tab:thr_ablation_ota1}
\vspace{1pt}
\scriptsize
\setlength{\tabcolsep}{1pt}
\renewcommand{\arraystretch}{1}
\resizebox{\columnwidth}{!}{%
\begin{tabular}{|l|c|r|c|r|}
\hline
\textbf{Metric} & \textbf{Iter. @15\%} & \textbf{Err. @15\%} & \textbf{Iter. @5\%} & \textbf{Err. @5\%} \\
\hline \hline
A$_{DC}$ & 2  & 2.57\% & 5 & 1.22\% \\ \hline
PM       & 8 & 0.17\% & 17 & 0.02\% \\ \hline
SR$^{+}$ & 12 & 10.86\% & 16 & 4.37\% \\ \hline
SR$^{-}$ & 11 & 10.54\% & 12 & 4.66\% \\ \hline
CMRR     & 2  & 9.61\% & 10 & 1.53\% \\ \hline
PSRR$^{+}$ & 7 & 9.44\% & 20 & 4.71\% \\ \hline
PSRR$^{-}$ & 9 & 4.37\% & 11 & 2.23\% \\ \hline
ICMR$_{min}$ & 8 & 5.63\% & 10 & 3.51\% \\ \hline
ICMR$_{max}$ & 1 & 13.68\% & 3 & 0.44\% \\ \hline
OCMR$_{min}$ & 3 & 5.05\% & 5 &  4.58\% \\ \hline
OCMR$_{max}$ & 2 & 3.15\% & 3 & 1.88\% \\ \hline
BW$_{3dB}$ & 12 & 4.43\% & 17 & 2.69\% \\ \hline
UGF      & 12 & 5.52\% & 14 & 3.49\% \\
\hline
\textbf{Max Iter. \ / Average Err.} 
& \textbf{12} & \textbf{6.54\%} 
& \textbf{20} & \textbf{2.67\%} \\
\hline
\end{tabular}%
}
\vspace{-5mm}
\end{table}

\noindent
\textbf{Sensitivity to convergence threshold:}
We repeat the repair loop for all five OTAs after tightening the error threshold from $15\%$ to $5\%$. The stricter threshold reduces the final average error to below $5\%$ for every topology, while keeping the maximum repair effort below 45 iterations. 
Table~\ref{tab:thr_ablation_ota1} shows detailed per-metric OTA-1 results as a representative case, where average final relative error decreases from $6.54\%$ to $2.67\%$.
This illustrates the expected tradeoff between stricter accuracy targets and increased repair effort.

\section{Conclusion}
\label{sec:conclusion}
\noindent
This paper presented \textsc{NEMESIS}, a multimodal LLM framework for generating OTA performance equations from SPICE netlists and schematics using SPICE-based verification and iterative repair. Across five OTA topologies, \textsc{NEMESIS} produced interpretable device-level equations whose predictions closely align with SPICE over assigned $g_m/I_D$ ranges. By providing physical circuit insight without manual derivation,
\textsc{NEMESIS} enables fast equation-based evaluation for repeated design exploration. However, (a)~it assumes that the topology is fixed; (b)~it does not replace final SPICE signoff. Extensions to large-signal nonlinearities, phase noise, high-frequency matching, and modeling for variable topologies in topology optimization are topics for future work.

\newpage
\balance
\bibliographystyle{misc/IEEEtran} 
\bibliography{bib/main.bib}
\end{document}